\documentclass[12pt]{article}
\usepackage{graphics,epsfig}
\newcounter{saveeqn}

\begin{document}

\begin{titlepage}\hfill HD-THEP-04-12\\[5 ex]

\begin{center}{\Large\bf Singular behavior of $^1P_1^{+-}$ quarkonium and 
positronium annihilation decays and relevance of relative energy} \\[5 ex]

{\large \bf Dieter Gromes}\\[3 ex]Institut f\"ur
Theoretische Physik der Universit\"at Heidelberg\\ Philosophenweg 16,
D-69120 Heidelberg \\ E - mail: d.gromes@thphys.uni-heidelberg.de \\
 \end{center} \vspace{2cm}

{\bf Abstract:} Using a four dimensional approach, we show that the 
singularities for small gluon momenta, which arise in the usual three 
dimensional treatment of the annihilation decay, disappear if all poles in 
the relative energy are taken into account correctly in the integration. We 
obtain an explicit formula for the decay width which involves a non locality 
originating from the kinetic energy. We  calculate not only the familiar 
logarithmic dependence on the binding energy, but also the constant to be 
added to the logarithm. For positronium this differs from an earlier result in 
the literature and leads to a modified life time. In QCD there is a
non abelian loop graph which contributes to the decay amplitude to the same 
order as the tree graph. Contrary to the original version, which contained
some errors, the constant turns out to be  large and negative.
 For reasonable parameters, this negative constant 
practically cancels the positive logarithm, so that the whole approach of 
expanding with respect to the binding energy breaks down.

 \vfill \centerline{March 2004, revised March 2005}

\end{titlepage}

\section{Introduction}

It is known for a long time that the usual non relativistic approach for 
calculating the decay 
width of the $^1P^{+-}_1$-state into three gluons, as well as the 
annihilation of $^3P_1^{++}$ into a gluon and light quark-antiquark pairs, 
has infrared divergences originating from soft gluon momenta. In the original 
paper by Barbieri, Gatto, and Remiddi \cite{BGR} the problem was overcome 
by introducing a fictituous width for the decay into a real and a virtual 
photon (or gluon). Their result was ($R_{21}(r)$ is the radial wave function)

\begin{equation} 
\Gamma (^1P_1^{+-} \rightarrow \mbox{3 gluons})  =  
 \frac{5}{18}\frac{8}{\pi}\frac{\alpha _s^3}{m^4}|R'_{21}(0)|^2 
 \bigg[ \ln  \frac{m}{|E|} + const. \bigg]. 
\end {equation}
Actually only the singular 
logarithmic dependence on the binding energy could thus be calculated, while 
the constant above was simply ignored. This is a shortcoming because such a 
constant, even if not particularly large, could be quite important since the 
logarithm is not large in QCD and even only about 12.6 in QED.

In a series of papers  Bodwin, Braaten, Lepage \cite{BBL} and collaborators
clarified some aspects of the problem by applying the methods of non 
relativistic QCD (NRQCD) \cite{NRQCD}. The quarkonium state is not only made 
up of quark and antiquark $(Q\bar{Q})$ but contains also Fock state 
components with quark-antiquark-glue $(Q\bar{Q}g)$ etc. Although this 
admixture of glue 
is suppressed  with respect to the dominant  $Q\bar{Q}$-state, it contributes 
to the decay with the same order. The reason is that the quark-antiquark 
annihilation in the $Q\bar{Q}$-state is also suppressed, because the 
$Q\bar{Q}$ wave function at the origin vanishes in the P-state. The 
$Q\bar{Q}$ in the $Q\bar{Q}g$-componet, on the other side, can be in an 
S-state, thus compensating the suppression of the Fock state component. The 
divergences found previously are canceled and the part with the logaritmic 
dependence on the binding energy can be calculated, the constant which 
accompanies the logarithm remains, however, unknown.

In a further paper on the subject the present author \cite{Gr} calculated the 
glue 
content in heavy quarkonia in order to obtain more definite statements. 
Although this calculation was succesful and gave reasonable results, it 
turned out that it was, at least up to now, not helpful for calculating the 
decay width. 

In the present work we go back to a more fundamental formalism. It involves 
the four dimensional T-matrix for the annihilation of the quark-antiquark 
pair into gluons, as well as the four dimensional Bethe Salpeter wave 
function, taken in the approximation coming from the static Salpeter 
equation. Both quantities depend on the relative energy $p^0 = (p_1-p_2)^0/2$ 
of the quark-antiquark pair. The usual non relativistic approach is 
equivalent to considering only the poles of the BS wave function when 
integrating over $p^0$. The $p^0$-dependence of the T-matrix is, however, 
ignored by fixing the quarks on mass shell there. As will be discussed in 
the following, this procedure is legitimate in most cases, like decays of 
S-wave states, or decays into two real gluons. It fails, however, for the 
decay of the $^1P^{+-}_1$-state into three gluons if one of the gluons is 
soft. 

The contributions from the poles in the T-matrix are only of relevance if one 
of the gluons is soft. In this region, however, they are crucial, because they 
cancel the singularities in the leading term. Thus the correct consideration 
of all poles gives an expression which is free of divergences. Another 
phenomenon which was already found in \cite{Gr} arises in the soft region: A 
non abelian graph, where a soft transversal gluon splits into two Coulomb 
gluons, while the latter couple to quark and antiquark, contributes to the 
same order as the tree graph. The reason is that the direct coupling of a 
soft transversal gluon to a non relativistic quark is also suppressed. 

Our result is not directly expressible by the derivative of the radial wave 
function at the origin, because it contains expressions of the form  
$k + |E| + {\bf p}^2/m$ in the denominator, with $k$ a gluon energy, $E$ the 
binding energy, and ${\bf p}^2/m$ the kinetic energy. The presence of the 
kinetic energy introduces a non locality. For a Coulomb potential one can 
carry through the calculation  analytically till the end. We agree with the 
formulae in the literature as far as  the logarithmic dependence on the 
binding energy is concerned. Beyond this we can calculate the constant term 
to be added to the logarithm. For the decay of the singlet P-state of 
positronium into three photons (where the non abelian contribution is of 
course absent) our constant differs from the one given by 
Alekseev \cite{alek}. In QCD the resulting constant is large 
and negative for reasonable parameters, and practically outweighs the 
positive logarithm. This means 
that the whole approach breaks down unless $\alpha _s$ is very small.

In our approach no $Q\bar{Q}g$ component in the wave function shows up at 
all. All contributions come from the four dimensional $Q\bar{Q}$ wave 
function alone. Nevertheless one may reinterprete our result in 
terms of the non relativistic picture and the presence of a quark-antiquark 
glue component in the wave function. 

In sect. 2 we discuss the situation and derive the general formalism. In 
sect. 3 we specialize to a Coulomb wave function and present an analytic 
formula. In sect. 4 we discuss the experimental status and future applications.

\setcounter{equation}{0}\addtocounter{saveeqn}{1}%

\section{Four dimensional versus three dimensional approach}

Consider an annihilation decay of a quarkonium state in the rest frame into  
three (or analogous into two) gluons. In the familiar three dimensional 
approach one proceeds as follows. The quark energies $p_1^0$ and $p_2^0$ in 
the T-matrix for $Q \bar{Q} \rightarrow$ gluons are taken 
on mass shell and the quark momenta smeared with the Schr\"odinger wave 
function. The amplitude, with the gluon momenta defined as in Fig. 1, then 
becomes

\begin{equation} <{\bf k}_1^a{\bf k}_2^b{\bf k}_3^c|K> = 
\frac{i(2\pi)^4\delta^{(4)}(k_1+k_2+k_3-K)}{\sqrt{(2\pi )^3m}}
\int T({\bf p})  \tilde{\psi }({\bf p})d^3p,
\end{equation} 
with $p^\mu = (p_1 - p_2)^\mu/2$ the relative momentum. The dependence of 
the T-matrix on the gluon momenta ${\bf k}_1,{\bf k}_2,{\bf k}_3$,  
polarization vectors $\epsilon _1,\epsilon _2,\epsilon _3$, and color 
indices $a,b,c$ has been suppressed. We used the standard covariant 
normalization of states and relativistic normalization of $T$, the 
Schr\"odinger wave function is normalized to 1.
The  T-matrix for quark-antiquark into gluons is a slowly varying function 
of ${\bf p}$, with a scale set by 
the quark mass $m$, while the wave function $\tilde{\psi }({\bf p})$ is 
non 
relativistic, i.e. dominated by soft momenta of the order of 
$\alpha _s m$. 
Therefore one can expand the T-matrix around ${\bf p}=0$. For S-states it 
is 
sufficient to consider the leading term $T(0)$, the 
remaining integral gives the wave function in position space at the 
origin, 

\begin{equation} T(0) \int \tilde{\psi }({\bf p})d^3 p = T(0) 
(2\pi)^{3/2} \psi (0). 
\end{equation} 
For P-states (with magnetic quantum number $m$) one has to expand up to 
first order in ${\bf p}$, resulting in 

\begin{eqnarray} \lefteqn{[(\nabla _{\bf p})^nT({\bf p})]\Big|_{{\bf p}=0} 
\int p^n \tilde{\psi }
({\bf p})d^3 p  = } \nonumber\\& &
   -i [(\nabla _{\bf p})^nT({\bf p})]\Big|_{{\bf p}=0}
Y_{1m}({\bf e}_{(n)}) (2\pi)^{3/2}R'(0),
\end{eqnarray} 
with $R(r)$ the radial wave function.

For decays into two gluons, which give the leading 
contribution for states with positive charge conjugation, i.e. $^1S^{-+}_0$ 
or $^3P_J^{++}$ for $J=0,2$ ($^3P_1^{++}$ cannot decay into two real gluons) 
there are no problems. Both gluons have to be hard from kinematical reasons. 
Therefore the denominator of the quark propagator in $T({\bf p})$ cannot 
vanish and no singularities can arise.

Let us next discuss the three gluon decay shown in Fig. 1 
which is the leading contribution for annihilation decays of states with 
negative charge conjugation, i.e. $^3S^{--}_1$ or $^1P^{+-}_1$. 

\begin{figure}[htb]
\begin{center}
\epsfysize 4cm
\epsfbox{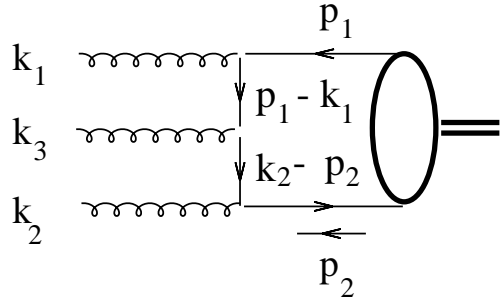}
\end{center}
\caption{ {\it The lowest order contribution to the three gluon decay.}}
\end{figure}    

With the quark energies taken on mass shell, and the quark momenta 
considered only up to first order, the denominators of the two quark 
propagators become (we always use $k_j \equiv |{\bf k}_j|$ in the following)

\begin{eqnarray} (p_1 - k_1)^2 -m^2 + i\epsilon & = & -2mk_1 
+ 2 {\bf p}_1 {\bf k}_1 +i\epsilon , \nonumber\\ 
(k_2 - p_2)^2 -m^2 + i\epsilon 
& = & -2mk_2 + 2 {\bf p}_2 {\bf k}_2 +i\epsilon. 
\end{eqnarray}
This is the well known reason for the possible appearence of divergences 
which can arise from small ${\bf k}_1$ or ${\bf k}_2$.

Let us now view the situation in the framework of a four dimensional 
treatment of the three gluon decay, focusing on the problematics for the 
P-state. The amplitude in Fig. 1 has the form

\begin{eqnarray} \lefteqn{<{\bf k}_1^a{\bf k}_2^b{\bf k}_3^c|K> = 
- \frac{(2\pi)^4\delta^{(4)}(k_1+k_2+k_3-K)}{\sqrt{(2\pi)^5 m}} \times }\\
& & \int Tr [T({\bf p},p^0)\Gamma ({\bf p})]  
\frac{  (|E|+{\bf p}^2/m) \;
\tilde{\psi }({\bf p}) \; dp^0 \; d^3p}{  
 (p^0 + |E|/2+ {\bf p}^2/2m  - i\epsilon) \; 
( p^0 - |E|/2- {\bf p}^2/2m  + i\epsilon)}.\nonumber
\end{eqnarray} 
As before, $T$ denotes the T-matrix for the annihilation of the 
quark-antiquark pair into gluons. $\Gamma ({\bf p})$ is the spin wave 
function of the bound state. For a singlet state it reads 
$\Gamma ({\bf p}) = \Lambda ({\bf p}_1) (\gamma ^5/\sqrt{2}) 
\Lambda ^c({\bf p}_2)$. 
Here $\Lambda ({\bf p}_k)$ are the projectors to positive energy states, in 
our case we need them only up to order ${\bf p}_k$ where they read 

\begin{equation} \Lambda ({\bf p}_1) = \frac{1+\gamma ^0}{2} 
+ \frac{\mbox{\boldmath $\alpha $} {\bf p}_1}{2m}, \mbox
{\quad and \quad } \Lambda ({\bf p}_2)^c = \frac{1-\gamma ^0}{2}
- \frac{\mbox{\boldmath $\alpha $} {\bf p}_2}{2m}. 
\end{equation} 
Finally $\tilde{\psi }({\bf p})$ is again the 
Schr\"odinger wave function, while $E$ is the binding energy. We used the 
well known $p^0$-dependence of the Bethe-Salpeter
wave function as it  arises from the static approximation to the 
 BS equation.

We first discuss the trace in the numerator. For the graph in Fig. 1 it is 

\begin{equation} Tr \Big\{\Gamma({\bf p}) \gamma ^{n_2}[\not{k}_2 
- \not{p}_2 +m]
\gamma ^{n_3}[(\not{p}_1-\not{k}_1)+m]\gamma ^{n_1}\Big\}. 
\end{equation}
This is only needed in zeroth order of ${\bf p}$ for S-states, and up to 
first order for P-states.
The zeroth order of the trace is proportional to $k_1k_2$ (if ultrasoft 
terms are dropped), therefore there is no trouble 
for the three gluon decay of S-states; the factors of the numerator cancel 
the singularities in ${\bf k}_1$ and ${\bf k}_2$ of the denominator. This is 
no longer the case for the terms of order ${\bf p}$.
The trace up to order ${\bf p}$, contracted with the polarization vectors,  
 is  

\begin{eqnarray}& & 2i \Big\{ \Big((k_1 + |E|/2 - p^0) (k_2 + |E|/2 + p^0) 
+ ({\bf k}_1 \cdot {\bf k}_2)\Big) 
[\epsilon _1 \times \epsilon _2] \cdot \epsilon _3 \nonumber\\
& & + (\epsilon _1 \cdot \epsilon _2) [ {\bf k}_1 \times {\bf k}_2] \cdot
\epsilon _3 
 + (\epsilon _1 \cdot {\bf k}_2) 
[\epsilon _3 \times \epsilon _2] \cdot {\bf k}_1
-  (\epsilon _2 \cdot {\bf k}_1) 
[\epsilon _3 \times \epsilon _1] \cdot {\bf k}_2\Big\}\nonumber\\
&  & + 4i\Big\{ (\epsilon _2 \cdot {\bf p}) [\epsilon _1 \times \epsilon _3] 
\cdot {\bf k}_1  
+  (\epsilon _1 \cdot {\bf p}) [\epsilon _2 \times \epsilon _3] 
\cdot {\bf k}_2  \Big\}.
\end{eqnarray}

We next have to perform the integration over $p^0$ in (2.5). 
The usual approach is equivalent to fixing the energies $p_1^0$ and $p_2^0$ 
on mass shell in the T-matrix, which means putting 
$p^0= |E|/2 + {\bf p}^2/(2m)$ in the first, and 
$p^0= -|E|/2 - {\bf p}^2/(2m)$ in the second quark propagator. Only the $p^0$ 
dependence from the Salpeter wave function is considered and the 
$p^0$-integration performed, thus ending up with the simple formula (2.1). 
In this way one has 
made three approximations:

\begin{itemize}
\item The values of $p^0$ in the T-matrix have been fixed by the on shell 
 prescription instead of using the residue at the pole of the wave function.
\item The contributions from the remaining poles have been neglected.
\item For P-states, where one has to expand up to order ${\bf p}$, the 
${\bf p}$-dependence of the projection operators $\Lambda $ has been ignored.
\end{itemize}

Let us come to the correct $p^0$-integration in the four dimensional formula 
(2.5).
The Salpeter wave function has a pole $\epsilon _w^+  $ in the upper half 
plane, and a pole $\epsilon _w^- $ in the lower half plane. There are 
four more poles from the 
two propagators in T, which we denote by $\epsilon _1^\pm  , $ and 
$\epsilon _2^\pm  $. Altogether the poles are

\begin{eqnarray}
\epsilon _w^\pm & = & 
\mp \; \Big(|E|/2 + {\bf p}^2/(2m)- i\epsilon \; \Big),\nonumber\\
\epsilon _1^\pm & = &  -m + |E|/2 + k_1 
\mp  \Big(\sqrt{m^2 + ({\bf k}_1 - {\bf p})^2}-i\epsilon \; \Big),\\
\epsilon _2^\pm & = &  \;\:\, m - |E|/2 - k_2 
\mp \Big(\sqrt{m^2 + ({\bf k}_2 + {\bf p})^2}-i\epsilon \; \Big). \nonumber
\end{eqnarray}

We have to calculate the following integral where the first four factors come 
from the quark propagators in the T-matrix, the last two from the four 
dimensional wave function:

\begin{equation}I = -\frac{1}{2\pi i}
 \int _{-\infty}^\infty\frac{(|E|+{\bf p}^2/m) \; dp^0}{ (p^0-\epsilon _2 ^+) 
(p^0-\epsilon _2^-  )(p^0-\epsilon _1 ^+ ) (p^0-\epsilon _1^-  )
(p^0-\epsilon _w ^+  ) (p^0-\epsilon _w^-  )}.
\end{equation}
We perform the $p^0$-integration by closing the integration contour in the 
upper half plane (which, temporarily, introduces an apparent asymmetry 
between quark and antiquark).
There are three contributions originating from the three poles at 
$\epsilon _w^+ , \; \epsilon_1^+ ,  \; \epsilon_2^+   $: 

\begin{equation} I = -(|E|+{\bf p}^2/m) \bigg(\frac{1}{D_w} + \frac{1}{D_1} 
+ \frac{1}{D_2}\bigg).
\end{equation}

The denominators which appear here are

\begin{equation} D_w =  {\prod_n}' (\epsilon_w^+ - \epsilon_n),\quad 
D_1 = {\prod_n}' (\epsilon_1^+ - \epsilon_n),\quad 
D_2 = {\prod_n}' (\epsilon_2^+ - \epsilon_n),  
\end{equation}
where the products  ${\prod_n}'$ run over the five remaining poles.

If all three gluons are hard, the contribution $1/D_w$ from the pole 
$\epsilon _w^+$ is dominant because $D_w$ contains the small ultrasoft 
difference $\epsilon _w^+ - \epsilon _w^- = -(|E| + {\bf p}^2/m)$. This 
cancels the corresponding factor in front of (2.11). All the other 
differences appearing in the three denominators are hard for hard gluons, 
therefore the contributions $1/D_1$ and $1/D_2$ are suppressed compared to the 
leading term $1/D_w$ by a relative factor of the order 
$(|E|/m+{\bf p}^2/m^2)$. The momentum $p_2$ is 
on mass shell on the pole $\epsilon ^+_w$, while $p_1$ is slightly off shell. 
One may, however, also put $p_1$ on mass shell in $D_w$, which corresponds to 
replacing $\epsilon _w^+ - \epsilon _1^{\pm}$ by $\epsilon _w^- 
- \epsilon _1^{\pm}$. This  again only causes a relative error of order 
$|E|/m$ and leads back to the old formula from the three dimensional treatment.

The situation is more subtle if one of the gluons is soft or ultrasoft. The 
other two are then, of course, hard. 

We start with the case where ${\bf k}_3 \rightarrow 0$, i.e. when the quark 
propagators in the T-matrix become identical and create a double pole. Both 
$D_1$ and $D_2$ become singular for ${\bf k}_3 \rightarrow 0$ which is due to 
the factor 

\begin{equation}\epsilon _1^+ - \epsilon _2^+ = -k_3 - \sqrt{m^2 + ({\bf k}_1 
- {\bf p})^2} + \sqrt{m^2 + ({\bf k}_2 + {\bf p})^2}
\end{equation} 
(we used $k_1 + k_2 + k_3 = 2m - |E|)$. This factor enters, however, with 
opposite signs in $D_1$ and $D_2$, while all the other factors become 
identical for ${\bf k}_3 \rightarrow 0$. Therefore there is no singularity 
for  ${\bf k}_3 \rightarrow 0$ in the sum.

For ${\bf k}_1 \rightarrow 0$ the first propagator gets near to the mass 
shell. This behavior shows up only in $D_w$ and is due to the difference

\begin{eqnarray}\epsilon _w^+ - \epsilon _1^- & = & -|E| - {\bf p}^2/(2m) 
+ m-k_1 - \sqrt{m^2 + ({\bf k}_1 - {\bf p})^2} \nonumber\\
& \rightarrow &
- (k_1 + |E| + {\bf p}^2/m - ({\bf k}_1 \cdot {\bf p})/m) 
\mbox{ for } {\bf k}_1 \rightarrow 0.
\end{eqnarray}
Finally we look at the limit ${\bf k}_2 \rightarrow 0$, where the second 
quark propagator becomes on shell. In $D_w$ there is a difference which 
becomes small, namely

\begin{eqnarray}\epsilon _w^+ - \epsilon _2^+ & = &  - {\bf p}^2/(2m) 
- m+k_2 + \sqrt{m^2 + ({\bf k}_2 + {\bf p})^2} \nonumber\\
& \rightarrow &
 k_2  + ({\bf k}_2 \cdot {\bf p})/m \mbox{ for } {\bf k}_2  \rightarrow 0.
\end{eqnarray}
In $D_2$ there are two differences which become small, namely 
$\epsilon _2^+ - \epsilon _w^+ \rightarrow -(k_2 
+ ({\bf k}_2 \cdot {\bf p})/m) $, i.e. the negative of the term above, and 

\begin{eqnarray}\epsilon _2^+ - \epsilon _w^- & = & -|E| - {\bf p}^2/(2m) 
+ m-k_2 - \sqrt{m^2 + ({\bf k}_2 + {\bf p})^2} \nonumber\\
& \rightarrow &
- (k_2 + |E| + {\bf p}^2/m + ({\bf k}_2 \cdot {\bf p})/m ) 
\mbox{ for } {\bf k}_2 \rightarrow 0.
\end{eqnarray}
In the sum the two poles at $ k_2 + ({\bf k}_2 \cdot {\bf p})/m= 0$ cancel, 
while the ultrasoft term

$-(k_2 + |E| + {\bf p}^2/m + ({\bf k}_2 \cdot {\bf p})/m)$ remains. 

Summarizing we found the following result: The contributions of $1/D_1$ and 
$ 1/D_2$ are negligible if all gluons are hard. They are also negligible for 
${\bf k}_3 \rightarrow 0$ because the poles cancel in the sum, as well as for 
${\bf k}_1 \rightarrow 0$, because they are finite in the latter limit. For 
${\bf k}_2 \rightarrow 0$ finally, the poles at ${\bf k}_2 =0$ in $1/D_w$ and 
$1/D_2$ cancel, while the pole (2.16) in $1/D_2$ survives. In total one thus 
obtains

\begin{equation} \frac{1}{D_w} + \frac{1}{D_1} + \frac{1}{D_2} = 
\frac{1+O(|E|/m+ {\bf p}^2/m^2)}{D},
\end{equation}
where $D$ is obtained from $D_w$ by replacing 
$\epsilon _w^+ - \epsilon _2^+$ by $\epsilon _w^- - \epsilon _2^+$. 
We may also replace 
$\epsilon _w^+ - \epsilon _2^-$ by $\epsilon _w^- - \epsilon _2^-$ because 
this difference is always hard. This restores the manifest symmetry between 
quark and antiquark. 

The spatial momentum ${\bf p}$ is soft in the bound state described by the 
wave function $\tilde{\psi }({\bf p})$. Therefore one can put ${\bf p}=0$ 
(as well as $E=0$) in all terms where ${\bf p}$-dependent terms are added to 
hard terms. 
In the two critical factors, i.e. in those which become small for small 
${\bf k}_1$ or ${\bf k}_2$ respectively, one can use the limits in the second 
lines of (2.14) and (2.16). The terms 
$({\bf k}_1 \cdot {\bf p})/m$ and $({\bf k}_2 \cdot {\bf p})/m$  
are treated in first order. This finally leads 
to the following simple form for the integral $I$ in (2.10):

\begin{equation} 
I =  \frac{1}{4m^2} \; 
\frac{1 + \frac{({\bf k}_1 \cdot {\bf p})}{m (k_1+|E| + {\bf p}^2/m)} 
- \frac{({\bf k}_2 \cdot {\bf p})}{m (k_2+|E| + {\bf p}^2/m)}
+O(|E|/m + {\bf p}^2/m^2)}{(k_1+|E| 
+ {\bf p}^2/m) \; (k_2+|E| + {\bf p}^2/m)}. 
\end{equation}
The apparent singularities for 
${\bf k}_1 \rightarrow 0$ and ${\bf k}_2 \rightarrow 0$, which showed up in 
the  three dimensional approach have turned out to be spurious. They 
disappear if all the poles in the relative energy $p^0$ are considered in 
the $p^0$-integration. For $j=1,2$ there remain, however, denominators of the 
form $k_j + |E| + {\bf p}^2/m$  which become small for small $k_j$ and 
enforce to keep a finite binding energy in order to avoid a divergence. 
Because the binding energy $E$ and the kinetic energy ${\bf p}^2/m$ are of 
the same order, one must, however, also keep the term ${\bf p}^2/m$ for 
consistency. This leads to a non local dependence upon the wave function.

We have to multiply (2.18) with the trace (2.8) and the remaining 
contributions from the expression (2.5). If terms contain e.g. a $k_1^n$  
in the numerator, one can replace $k_1+|E|+{\bf p}^2/m \rightarrow k_1$ in 
the denominators of (2.18). This finally leads to

\begin{eqnarray}\lefteqn{ 
<{\bf k}_1^a{\bf k}_2^b{\bf k}_3^c|K>_{tree} 
 = \bigg( \frac{d^{abc}}{4\sqrt{N}} \bigg) \; 
i(2\pi)^4\delta^{(4)}(k_1+k_2+k_3-K) \times }
\nonumber\\
& & \frac{g^3}{\pi^{3/2} m^{3/2}} \int F({\bf p})  \tilde{\psi}
({\bf p}) 
\; d^3p + \mbox{ permutations,}
\end{eqnarray} 

with

\begin{eqnarray} F({\bf p}) & = & \frac{ 
[\epsilon _1 \times \epsilon _3] \cdot {\bf k}_1 
(\epsilon _2\cdot {\bf p}) }{k_1 (k_2 + |E| + {\bf p}^2/m)} 
 +  \frac{1}{4 m k_1 k_2} 
\bigg(\frac{({\bf k}_1 \cdot {\bf p})}{k_1} - 
      \frac{({\bf k}_2 \cdot {\bf p})}{k_2}\bigg)  \times \nonumber\\
& & \bigg[ \Big(k_1 k_2 + ({\bf k}_1 \cdot {\bf k}_2)\Big) 
[\epsilon _1 \times \epsilon _2] \cdot \epsilon _3 
+ (\epsilon _1 \cdot \epsilon _2) [ {\bf k}_1 \times {\bf k}_2] \cdot
\epsilon _3 + 2 (\epsilon _1 \cdot {\bf k}_2) 
[\epsilon _3 \times \epsilon _2] \cdot {\bf k}_1 \bigg]. \nonumber
\end{eqnarray}

The permutations are the remaining five permutations of the three gluons. 
The factor $(d^{abc}/(4\sqrt{N}))$ in front is the color factor, with $N=3$ 
the number of colors. For the decay of the singlet P-state of positronium 
into three gammas it has to be replaced by 1, while $g$ is replaced by $e$. 
The first term in $F({\bf p})$ originates from the product of the order 
${\bf p}$ term of the trace with the order 1 term of the denominator,
the remaining contributions come from the order 1 term of the trace, 
multiplied with the order ${\bf p}$ term of the denominator.

In QED (2.19) would in fact be the leading contribution to the decay 
amplitude. In QCD we encounter, however, the same phenomenon which we found 
already in \cite{Gr} when calculating the admixture of the 
quark-antiquark-glue component to the wave function. For the case that 
${\bf k}_2$, say, is ultrasoft, the loop graph in Fig. 2, where the dashed 
lines denote Coulomb gluons, contributes to the same order as the tree graph 
in Fig. 1 (although it will not lead to a logarithmic dependence upon the 
binding energy). 

\begin{figure}[htb]
\begin{center}
\epsfysize 4cm
\epsfbox{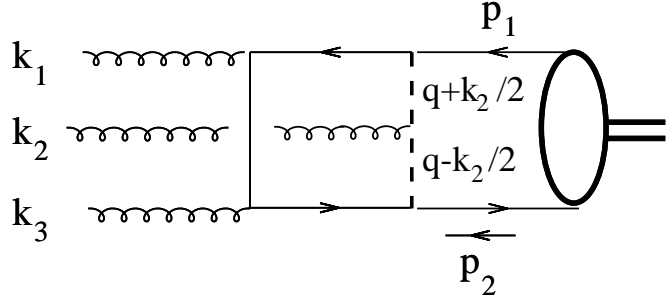}
\end{center}
\caption{ {\it The loop graph for the three gluon decay. Note that the 
indices of the gluons have been chosen such that the dominant contribution 
comes from $k_2$ soft.}}
\end{figure}    

The reason is that the tree graph is suppressed by the coupling of a 
transverse gluon to a non relativistic quark. The loop graph, on the other 
hand, is suppressed by two more coupling constants, but this is compensated 
because the coupling to the quarks via Coulomb gluons is not suppressed. 
Other loop graphs are not relevant, because they don't have a sufficient 
number of denominators which can become small simultaneously. The same is 
true for higher order corrections in the wave 
function, e.g. exchange of a transverse gluon. Also the graph in Fig. 2 is 
only relevant in the region of ultrasoft ${\bf k}_2$. This simplifies the 
calculation considerably.

If one first integrates over $q^0$ in Fig. 2 one finds a similar situation as 
before. Spurious poles in $k_1$ and $k_3$ which show up at some residues 
cancel in the sum. Only the variable $k_2$ appears in a denominator which 
becomes small but finite for ultrasoft $k_2$. After having performed the 
$q^0$-integration, the whole $p^0$-dependence is in the BS wave function and 
the $p^0$-integration can be immediately performed. Finally one can do the 
integration over $d^3q$ which is simplified by the fact that one can drop 
the $k_2$ in the Coulomb propagators. The resulting expression is 

\begin{eqnarray}\lefteqn{ <{\bf k}_1^a{\bf k}_2^b{\bf k}_3^c|K>_{loop}  = 
- \bigg( \frac{d^{abc}}{4\sqrt{N}} \bigg) \; 
i(2\pi)^4\delta^{(4)}(k_1+k_2+k_3-K) \times }\\
& & \frac{ g^5 N
[\epsilon _1 \times \epsilon _3] \cdot {\bf k}_1
}{16 \pi ^{5/2} \sqrt{m} } 
 \int \frac{(\epsilon _2\cdot {\bf p})}{p^3} f
\bigg( \frac{\sqrt{m (k_2+|E|)}}{p}\bigg) \; \tilde{\psi}({\bf p})\; d^3p 
\nonumber\\
& & + \mbox{ permutations,}\nonumber
\end{eqnarray} 
with

\begin{equation} f(x) = \frac{\pi}{2} - \frac{x}{1+x^2} -\arctan x.
\end{equation}

\setcounter{equation}{0}\addtocounter{saveeqn}{1}%

\section{Three gluon (or 3 $\gamma $) annihilation of the $^1P_1^{+-}$ state}

Due to the non local dependence on the momentum ${\bf p}$ in (2.19), (2.20) 
one cannot 
simply express the result by the derivative of the wave function at the 
origin. Instead one has to perform the $p$-integration explicitly. We carry 
this through for the case of the lowest Coulomb P-state wave function which 
reads (we drop an irrelevant factor $i$)

\begin{equation}
\tilde{\psi }({\bf p}) = \frac{8\sqrt{2}\sqrt{m |E|}}{\sqrt{\pi}} R'_{21}(0) 
\frac{|{\bf p}|}{(|{\bf p}|^2+m |E|)^3} Y_{1m}({\bf \hat{p}}),
\end{equation}
with $|E| \equiv |E_2| = (C_F\alpha _s)^2m/16$ and $C_F = (N^2-1)/(2N) 
= 4/3$. To make contact with the literature we have split off the derivative 
of the wave function at the origin, $R'_{21}(0) = 2(m|E|)^{5/4}/\sqrt{3}$. 
The $p$-integration in (2.19) and (2.20) can now be performed and we obtain

\begin{eqnarray}
 <{\bf k}_1^a{\bf k}_2^b{\bf k}_3^c|K>_{Coulomb} \; & = &  
\bigg( \frac{d^{abc}}{4\sqrt{3}} \bigg) \; i(2\pi)^4\delta^{(4)}
(k_1+k_2+k_3-K)
R'_{21}(0)  
\times \nonumber\\
& &  [g^3 f_{tree}(k_1,k_2) +g^5  f_{loop}(k_1,k_2)] + 
\mbox{ permutations.}
\end{eqnarray} 
The functions $f_{tree}(k_1,k_2)$ and $f_{loop}(k_1,k_2)$  arise from the 
tree graph and the loop graph, respectively and read

\begin{eqnarray}\lefteqn{f_{tree}(k_1,k_2)  = }\nonumber\\
& & \frac{2\sqrt{2}}{3m^{3/2}k_1k_2^3} \;
[\epsilon _1 \times \epsilon _3] \cdot {\bf k}_1 Y_{1m}
(\epsilon _2) \; 
\Big( 8E^2+12|E|k_2+3k_2^2-8\sqrt{|E|}(|E|+k_2)^{3/2} \Big)
\nonumber\\
& + & \frac{1}{\sqrt{2} m^{5/2} k_1k_2} 
\Big[ Y_{1m}(\frac{{\bf k}_1}{k_1})- Y_{1m}(\frac{{\bf k}_2}{k_2})
\Big] 
\times\nonumber\\
& & \bigg[ \Big(k_1 k_2 + ({\bf k}_1 \cdot {\bf k}_2)\Big) 
[\epsilon _1 \times \epsilon _2] \epsilon _3   
+ (\epsilon _1 \cdot \epsilon _2) [ {\bf k}_1 \times {\bf k}_2] 
\epsilon _3 + 2 (\epsilon _1 \cdot {\bf k}_2) 
[\epsilon _3 \times \epsilon _2] \cdot {\bf k}_1 \bigg],
\nonumber
\end{eqnarray}
\begin{eqnarray}
f_{loop}(k_1,k_2) & = & - \; \frac{1}{2 \sqrt{2} \pi m^2} \; 
[\epsilon _1 \times \epsilon _3] \cdot {\bf k}_1 Y_{1m}
(\epsilon _2) \;
\frac{1}{\Big( \sqrt{|E|} + 
\sqrt{k_2+|E|}\Big) ^3}.
\end{eqnarray}
In the calculation of the width we use the gluon energies $k_1 $ and $k_2 $ 
as independent variables in the phase space integral, the region of 
integration is given by the triangle 

\begin{equation}
0  \leq  k_1,k_2 \leq m-|E|/2 \approx m, \quad
 k_1+k_2  \geq  m-|E|/2 \approx m.
\end{equation} 
As indicated, one can neglect the ultrasoft binding energy in the boundaries.

One next has to square the T-matrix in (3.2), perform the summation over 
gluon polarizations, and integrate over $k_1$ and $k_2$. Of course one can 
omit the permutations in one factor and, instead, multiply by a factor 6. 
The integration over $k_1$ gives a factor $k_2$ from the boundaries of phase 
space. Therefore only those terms are sensitive to the binding energy $E$ 
where both factors become small for small $k_2$ (see the structure of 
(2.19)). For the tree term this is only the case in two permutations, namely 
the identity and the exchange $k_1 \leftrightarrow k_3$, in the remainig 
four one may drop $E$  and perform both integrations analytically. In the 
two  potentially ''dangerous'' terms, one has to keep the binding 
energy. The $k_2$-integration can be performed analytically (with the help of 
Mathematica). The remaining $k_1$-dependent function can then be split in two 
terms. The first term can be integrated analytically and contains the 
logarithmic dependence on $E$, while in the second term one can drop $E$ and 
subsequently also perform the $k_1$-integration analytically. In the square 
of the non abelian term, as well as in the mixed term, only the identity and 
the permutation $k_1 \leftrightarrow k_3$ contribute. One may put $E=0$ 
there, no logarithmic singularity arises from these terms. 

For positronium, where the non abelian contribution is absent, the result for 
the decay width into three gammas finally becomes

\begin{eqnarray}\lefteqn{ 
\Gamma (^1P_1^{+-} \rightarrow 3 \gamma) =}\nonumber\\
  &  &
 \frac{8}{\pi}\frac{\alpha ^3}{m^4}|R'_{21}(0)|^2  \bigg[ \ln  
\frac{m}{|E|} 
- (\frac{25}{9} + \frac{\pi ^2}{64} + 2 \ln 2)  
- (\frac{3 \pi ^2}{32} - \frac{3}{4})
 + (\frac{49 \pi ^2}{192}  - \frac{5}{2}) \bigg]  \nonumber\\
  & = &   
 \frac{\alpha ^8  m}{96 \pi} 
 \bigg[ \ln  \frac{16}{\alpha^2} 
- (\frac{25}{9} + \frac{\pi ^2}{64} + 2 \ln 2)  
 - (\frac{3 \pi ^2}{32} - \frac{3}{4})
 + (\frac{49 \pi ^2}{192}  - \frac{5}{2}) \bigg]. 
\end {eqnarray}
The width for the quarkonium decay into three gluons becomes 
(5/18 is the color factor)

\begin{eqnarray}\lefteqn{ 
\Gamma (^1P_1^{+-} \rightarrow \mbox{3 gluons})  =  
 \frac{5}{18}\frac{8}{\pi}\frac{\alpha _s^3}{m^4}|R'_{21}(0)|^2 
\times } 
\nonumber\\
& & \bigg[ \ln  \frac{m}{|E|} - (\frac{25}{9} + \frac{\pi ^2}{64} 
+ 2 \ln 2)
- (\frac{3 \pi ^2}{32} - \frac{3}{4})
 + (\frac{49 \pi ^2}{192}  - \frac{5}{2}) - \frac{9}{8} + \frac{9}{64}\bigg]. 
\end {eqnarray}
Let us first discuss the positronium case (3.5). Comparing with the formula 
of Alekseev \cite{alek}  which reads

\begin{equation} 
\Gamma (^1P_1^{+-} \rightarrow 3 \gamma)_{Alekseev}  =  
 \frac{\alpha ^8  m}{96 \pi} 
 \ln  \frac{32}{\alpha^2}, 
\end{equation} 
we see that the logaritmic term $\ln 16/\alpha ^2$ coincides, but the 
constant differs. Instead 
of $\ln 2 = 0.693$ in Alekseev's formula we have 
$ - (\frac{25}{9} + \frac{\pi ^2}{64} + 2 \ln 2)  
- (\frac{3 \pi ^2}{32} - \frac{3}{4})
 + (\frac{49 \pi ^2}{192}  - \frac{5}{2}) = -4.47475$.
The life time 
according to our formula becomes $\tau = 0.00594$ s, 
which is a factor of 1.6  larger than the life time which is derived from 
(3.7). 

There is also a connection to another interesting  problem of QED. 
Recently Manohar and Ruiz-Femen\'{\i}a \cite{MRF} reanalyzed the low energy 
photon spectrum of the three photon decay of orthopositronium in the 
framework of NRQCD. This work was motivated by the observation of Pestieau 
and Smith \cite{PS}, that the spectrum, as derived by Ore and Powell 
\cite{OP} long ago, violates Low's theorem \cite{Low}. The authors 
found agreement with the Ore Powell formula for the case that all photons are 
hard, but obtained deviations if one photon becomes soft. The low energy 
photon spectrum (the energies of the remaining two photons are integrated) 
does not behave like the Ore Powell spectrum but is much softer.

One can try to understand the origin of such a suppression in our approach. 
An analogous calculation for the orthopositronium $^3S_1^{--}$ would show that 
the $k$ has been replaced by $k + |E| + {\bf p}^2/m$ in some of the 
denominators, i.e. for $k \rightarrow 0$ there is a suppression factor. This
is, however, not sufficient to give the correct result. 
The point is, that for ultrasoft $k_j$, electric or magnetic radiative 
transitions to the states $^3P_{0,2}^{++}$ or $^1S_0^{-+}$, which 
subsequently annihilate into two photons, become relevant. To take into 
account these bound states correctly, one would need the exchange of a whole 
ladder. For this special problem  the methods of non relativistic QED are 
certainly more appropriate (I thank P. Ruiz-Femen\'{\i}a for an enlightening 
correspondence).

We next discuss our result (3.6) for the quarkonium decay into three gluons.
As far as the logarithmic term is concerned it coincides with the formula 
given by Barbieri, Gatto, and Remiddi \cite{BGR}. Beyond this we were able to 
calculate the associated constant term to be added. 
The five contributions arise from the square of the expansion (in first 
order with respect to ${\bf p}$) of the numerator of 
the tree graph, the mixed term between numerator and denominator, 
the square 
of the expansion of the denominator, the mixed term between tree 
graph and 
loop graph (only the expansion of the numerator contributes), and 
the square 
of the loop graph. Numerically they are 

$-4.31828 - 0.17528 + 0.01881 - 1.12500 + 0.14062 = -5.45913.$

To get a feeling for the importance of the constant term let's use 
$|E|/m = (4 \alpha _s /3)^2/16$  and choose 
$\alpha _s = 0.2$ which  gives $m/|E| = 225$ and $\ln (m/|E|) = 
5.4161$. This 
shows that the negative constant outweighs the positive logarithm and 
formally leads to a negative widths! The whole approach breaks down 
and makes only sense if $m/|E|$ is much larger than the value above.
(In the original version there were some errors which led to a very small 
constant. The conclusions thus have changed drastically!)

Some remarks on the error are appropriate. It is a term of the order 
$\sqrt{|E|/m}$ in the brackets of (3.5) and (3.6). A consistent calculation 
of this correction would be extremely cumbersome, considering the many places 
where we have dropped the binding energy. It is, however, possible to 
calculate the correction coming from the last step, when performing the 
integration in the two permutations which lead to the logarithmic term. It 
results in the addition of a term $13.1 \sqrt{|E|/m} \approx 0.87$ in the 
bracket of (3.6). The correction is thus not dramatic, in spite of the quare 
root behavior and the large factor in front.

One might be tempted to simplify the whole procedure by replacing the non local
${\bf p}^2/m $-dependence in  (2.18), (2.19) by some average 
$\overline{{\bf p}^2/m}$. This would directly lead to a result involving 
$|R'(0)|^2$ for any wave function. In the non abelian loop graph there is a 
dependence on the variable $({\bf p} - {\bf q})^2$, and it is hard to find a 
reasonable approximation for this. Therefore we restrict the discussion to 
the abelian case. The result would be

\begin{equation}
 \Gamma (^1P_1^{+-} \rightarrow 3 \gamma)  \approx
\frac{\alpha ^8  m}{96 \pi} 
 \bigg[ \ln  \frac{16}{\alpha ^2} - \Big(\frac{5}{2} - \frac{7 \pi ^2}{48} 
+  \ln (1+\overline{{\bf p}^2/(m|E|)} \Big) \bigg]. 
\end {equation}
One could next use the virial theorem for the Coulomb potential and put 
$\overline{{\bf p}^2/m}=|E|$.  This would obviously lead to a constant 
different from that in (3.5). The width would be multiplied by a factor 
of 1.33 as compared to (3.5). Thus one obtains a bad approximation even in 
QED where the logarithmic term $\ln (1/\alpha ^2)$ dominates. This result 
shows that the nonlocality is essential. On the other hand it demonstrates 
the stability of the logarithmic term  $\sim \ln (1/\alpha ^2)$ against any 
reasonable approximation which respects the fact that ${\bf p}^2/m$ and $E$ 
are of the same order.

\setcounter{equation}{0}\addtocounter{saveeqn}{1}%

\section{Conclusions}

To our knowledge there are no data on the decay of the singlet P positronium 
state. For the $c\bar{c}$-system there is only a doubtful 
candidate \cite{hc} at 3526 MeV for the $^1P_1^{+-}$ state $h_c$, with width 
$<$ 1.1 MeV and no annihilation decays observed. For the $b\bar{b}$ system 
there is not even a candidate. This is unfortunate because these states would 
not only provide an excellent further test for the nature of the long range 
spin dependent potential, but also for our ideas about annihilation decays. 
To do this one has, however, to go beyond the approach of expanding with
respect to the binding energy. Our prediction for the annihilation width is 
a  modification of that given in \cite{BGR} and reads

\begin{equation}
\frac{\Gamma (^1P_1^{+-})}{\Gamma (^3P_0^{++})} =\frac{10}{27}
\frac{\alpha _s}{\pi} 
(\ln \frac{m}{|E|} - 5.459). \end{equation}
This shows again that the whole approach can only be trusted for 
very small binding energies.

From a purely theoretical point of view the annihilation decays under 
consideration are highly interesting and provided some puzzles in the past. 
We have shown here that a four dimensional treatment of the decay immediately 
leads to a result which is free of infrared diverences and which, 
furthermore, allows a definite calculation for any given wave function. 
Notwithstanding the fact that NRQCD has certainly greatly improved our 
understanding of low energy QCD and has led to a systematic and effective way 
for calculations, we believe that for some problems, like the one discussed 
here, a four dimensional treatment is more transparent and effective. 

We derived our result by using the $Q\bar{Q}$- wave function only. 
Nevertheless it is easy to make contact with the three dimensional approach. 
The contributions from the poles in the T-matrix can, alternatively, be 
interpreted as contributions from $Q\bar{Q}g$ components (with a soft gluon) 
in the wave 
function. The advantage in our approach is that we obtain this contribution 
immediately in an explicit form. Obviously these two ways of viewing the 
situation would also apply to decays into an arbitrary number of gluons.

The decay of the spin triplet state $^3P_1^{++}$ into a real gluon plus a 
virtual gluon which further decays into a light quark-antiquark pair shows 
the same infrared problems as the 
decay discussed here. Phenomenologically it is more interesting, because the 
charmonium state $\chi _{c1}(1P)$ is experimentally well established and 
theoretical results can be compared with the data. This will be undertaken 
in a forthcoming work.

\vspace{1em}
{\bf Acknowledgement:} I thank D. Melikhov for reading the manuscript and 
for valuable suggestions.

\end{document}